# A Two-Dimensional Hydrostatically Equilibrium Atmosphere of a Neutron Star with Given Differential Rotation

V. S. Imshennik* and K. V. Manukovskiĭ

*Institute of Theoretical and Experimental Physics, ul. Bol'shaya Cheremushkinskaya 25, Moscow, 117259 Russia*

**Abstract**—An analytic solution has been found in the Roche approximation for the axially symmetric structure of a hydrostatically equilibrium atmosphere of a neutron star produced by collapse. A hydrodynamic (quasi-one-dimensional) model for the collapse of a rotating iron core in a massive star gives rise to a heterogeneous rotating protoneutron star with an extended atmosphere composed of matter from the outer part of the iron core with differential rotation (Imshennik and Nadyozhin, 1992). The equation of state of a completely degenerate iron gas with an arbitrary degree of relativity is taken for the atmospheric matter. We construct a family of toroidal model atmospheres with total masses $M \approx 0.1$–$0.2 M_\odot$ and total angular momenta $J \approx (1$–$5.5) \times 10^{49}$ erg s, which are acceptable for the outer part of the collapsed iron core, in accordance with the hydrodynamic model, as a function of constant parameters $\omega_0$ and $r_0$ of the specified differential rotation law $\Omega = \omega_0 \exp[-(r\sin\theta)^2/r_0^2]$ in spherical coordinates. The assumed rotation law is also qualitatively consistent with the hydrodynamic model for the collapse of an iron core.

Key words: *pulsars, neutron stars and black holes; plasma astrophysics, hydrodynamics, and shock waves*

**1.** The gravitational collapse of an iron stellar core in the case of its sufficiently rapid initial rotation can produce a binary of neutron stars (NSs) (Imshennik 1992) and surrounding residual gaseous medium that does not become part of the rotating protoneutron star, which is generally unstable to fragmentation (Aksenov *et al.* 1995) and to conversion into a NS binary. The presence of a gaseous medium can result, first, from the heterogeneity of collapse, as revealed by hydrodynamic modeling (Imshennik and Nadyozhin 1992), for the outer part of the collapsing iron core, and, second, from subsequent protoneutron-star fragmentation, as was found in three-dimensional hydrodynamic models (Houser *et al.* 1994; Aksenov 1999); however, the latter models have failed to yield a NS binary and only give the ejection of a gaseous envelope. In any case, the gaseous medium is most likely an iron gas with a completely degenerate electron component, which is in hydrostatic equilibrium in the external gravitational field produced almost entirely by the more massive NS of the putative binary if its mass exceeds appreciably the total mass of the iron gas and the less massive NS. It should be noted that both NSs (particularly the more massive one)—the fragmentation products of the protoneutron star—lie in the immediate vicinity of the center of a vast iron-core cavity. Therefore, a satisfactory approximation for the problem of hydrostatic equilibrium of an iron gas is the Roche approximation with a central source for the gravitational field of the more massive NS with $M_p \approx (1.4$–$1.8)M_\odot$, which must be equal to the initial iron-core mass $M_{Fe} \approx 2.0 M_\odot$ minus the iron-gas mass $M_e \approx (0.1$–$0.2)M_\odot$ and the mass of the low-mass NS $M_{ns} \approx (0.1$–$0.4)M_\odot$. In the quasi-one-dimensional approximation with the centrifugal force averaged over angle $\Theta$ (in spherical coordinates), this problem was approximately solved by Imshennik and Zabrodina (1999). These authors showed the possibility of hydrostatic equilibrium for a gas mass of $\sim 0.1 M_\odot$ if the centrifugal forces (on the average) accounted for 90% of the gravitational forces of attraction with a central mass $M_p = 1.9 M_\odot$. The remainder of the repulsive forces are produced by electron gas pressure, with $\rho \propto r^{-3/2}$ and the external radius $r_{ex} \simeq 2.4 \times 10^8$ cm, which is almost half the initial radius $R_{Fe} \simeq 4.4 \times 10^8$ cm of a $2M_\odot$ iron core. In order to estimate the radial sizes of the atmosphere, $r_{ex}$ should be compared with the radius $r_{in}$ of the more massive NS, rather than with the binary orbital radius, if the inequality $M_p \gg M_{ns}$ is satisfied for all times of the binary evolution. Indeed, from the very outset of the binary existence, i.e., since fragmentation, the protoneutron-star separation into two unequal parts seems more likely. One of the arguments for this separation is the detection of an early neutrino signal from SN 1987A, which, unfortunately, was recorded by only one detector (LSD) 4.7 h before the main neutrino signal recorded by three detectors (IMB, KII, BST). If we, nevertheless, associate this early signal with fragmen-

---

tation and the main signal with the collapse of the more massive NS, then, according to our analytic model for the evolution of a NS binary with a total mass of $2M_\odot$ (Imshennik and Popov 1998), the initial mass ratio is $M_{ns}/M_p = 0.195$[1]. During the evolution of the binary orbit, the above ratio, after the low-mass NS fills the Roche lobe, decreases at the end of the evolution to $M_{ns}/M_p = 0.0526$, when $M_{ns} \simeq 0.1 M_\odot$ and $M_p \simeq 1.9 M_\odot$. Calculations of fragmentation in the above three-dimensional hydrodynamic model with the ejection of a low-mass envelope provide another argument for the separation into two unequal parts. Here, we take the radius of the more massive NS to be $r_{in} \simeq 6 \times 10^6$ cm, bearing in mind that the radius of the Lagrangian layer of a $1.8 M_\odot$ protoneutron star was found in the hydrodynamic model of Imshennik and Nadyozhin (1992) to be $3 \times 10^6$ cm (see also Imshennik 1995). The increase in the NS radius can be easily understood by taking into account the large contribution of centrifugal forces when the angular momentum is partially conserved during fragmentation. The adopted two-fold increase of this estimate was dictated by convenience of our subsequent computation of the atmospheric structure and is essentially of no importance for our results (see below). Thus, we disregard the mass of the low-mass NS, along with the mass of the atmosphere itself, as sufficiently small and very distant from the coordinate origin with a characteristic orbital radius $r_{orb} \sim 10^8$ cm (Imshennik and Popov 1998).

**2.** The basic equations for the hydrostatic equilibrium of a two-dimensional configuration with a central pointlike or spherically symmetric gravitating body of given mass $M_p$ and with a given rotation law $\Omega = \Omega(\tilde{\omega})$, where $\tilde{\omega}$ is the cylindrical radius, can be derived from the most general, unsteady-state three-dimensional hydrodynamic system with self-consistent "gravitation," for example, from Aksenov and Imshennik (1994). A given rotation law (with no dependence on the $z$ coordinate) for barotropic equations of state $P = P(\rho)$ is known to be necessary for the conditions of hydrostatic equilibrium to be satisfied (Tassoul 1978).

However, the equations of interest to us in the axisymmetric case can be easily derived. Each Lagrangian particle of matter is subjected to the radially directed gravitational force of the point source, $F_g = -GM_p/r^2$, and to the centrifugal force, $F_r = \vartheta_\varphi^2/\tilde{\omega}$, directed perpendicular to the rotation axis (taken as the $z$ axis), from which the angle $\theta$ of the spherical coordinate system is measured. In that case, an external force (per unit mass), which we write in the right-hand part of the relation

$$\frac{1}{\rho}\frac{\partial P}{\partial r} = F_r \cos\left(\frac{\pi}{2} - \theta\right) - \frac{GM_p}{r^2} = \frac{\vartheta_\varphi^2}{r} - \frac{GM_p}{r^2}, \quad (1)$$

acts in the radial direction and an external force, which is also written on the right,

$$\frac{1}{\rho r}\frac{\partial P}{\partial \theta} = F_r \cos\theta = \frac{\vartheta_\varphi^2}{r\sin\theta}\cos\theta = \frac{\vartheta_\varphi^2}{r}\cot\theta \quad (2)$$

acts in the meridional direction tangential to a circumference of radius $r$ centered at the coordinate origin.

The counterbalancing external forces of the $r$- and $\theta$-components of the pressure gradient are written in the left-hand parts of Eqs. (1) and (2). Thus, the system of equations of hydrostatic equilibrium for matter with a barotropic equation of state $P = P(\rho)$ takes explicit form in spherical coordinates if we substitute the relations for the azimuthal rotation velocity $\vartheta_\varphi = (r\sin\theta)\Omega$, where $\Omega = \Omega(r\sin\theta)$ is so far an arbitrary function of the only argument $r\sin\theta$ in Eqs. (1) and (2):

$$\frac{1}{\rho}\frac{\partial P}{\partial r} = r\Omega^2\sin^2\theta - \frac{GM_p}{r^2}, \quad (3)$$

$$\frac{1}{\rho}\frac{\partial P}{\partial \theta} = r^2\Omega^2\sin\theta\cos\theta. \quad (4)$$

Clearly, the matter density $\rho = \rho(r, \theta)$ can be determined from this system by additionally specifying the equation of state and boundary conditions. This is done below, but first we formulate some corollaries obtained during a more detailed derivation of the equations from the general system of unsteady-state, three-dimensional hydrodynamic equations in Aksenov and Imshennik (1994). This derivation is contained in Manukovskiĭ (1999). The main corollary is as follows: in the Roche approximation, the system of equations (3) and (4) is unique for a hydrostatically equilibrium ($\partial/\partial t = 0$), axially symmetric ($\partial/\partial\varphi = 0$) atmosphere composed of matter with a barotropic equation of state $P = P(\rho)$. In this case, it is, of course, implied that the specific internal energy $E$ is given by the thermodynamic relation $P = \rho^2 dE/d\rho$. Other corollaries are formulated as follows: (i) we first exclude a singular rotation law, $\Omega \propto \tilde{\omega}^{-2}$, at nonzero velocity components, $\vartheta_r \neq 0$ and $\vartheta_\theta \neq 0$, and (ii) then rule out the possibility of nonzero velocities; i.e., we assume that $\vartheta_r = \vartheta_\theta \equiv 0$. The latter corollary follows from such an unacceptable property of these velocity components in the steady-state, axisymmetric case: their resultant at any point in space is directed along the rotation axis, because the identity $\vartheta_r\sin\theta + \vartheta_\theta\cos\theta = 0$ holds. This all implies that the well known meridional matter circulations, which, in contrast to the above ones, have closed trajectories and require no external sources of matter (Mestel 1970), are not possible.

Below, we use the equation of state of a completely degenerate electron gas with an arbitrary degree of rel-

---

[1] The estimate is given for a circular orbit (with a zero eccentricity). For elliptic orbits, this ratio slightly increases, but this increase is generally modest as long as $e_0 \leq 0.8$ (Imshennik and Popov 1994).

ativity, i.e., we imply a nonzero entropy of matter composed of $^{56}$Fe, so the electron density is

$$n_e = \frac{26}{56} \frac{\rho}{m_u}. \tag{5}$$

In this case, we have for the pressure $P$ and the specific internal energy $E$ (see Landau and Lifshitz 1951)

$$P = C\left[\xi_F\left(\frac{2}{3}\xi_F^2 - 1\right)\sqrt{\xi_F^2 + 1} + \mathrm{Arsh}\xi_F\right], \tag{6}$$

$$E = \frac{C}{\rho}[\xi_F(2\xi_F^2 + 1)\sqrt{\xi_F^2 + 1} - \mathrm{Arsh}\xi_F], \tag{7}$$

where the dimensionless Fermi momentum (in units of $m_e c$) is

$$\xi_F = B\rho^{1/3} \tag{8}$$

and the constants $B$ and $C$ have the following known values:

$$B = \frac{2\pi\hbar}{m_e c}\left(\frac{39}{224\pi m_u}\right)^{1/3} = 7.792 \times 10^{-3} \text{ cm/g}^{1/3},$$

$$C = \pi m_e c^2 \left(\frac{m_e c}{2\pi\hbar}\right)^3 = 1.801 \times 10^{23} \text{ g(s}^2\text{ cm)}^{-1}.$$

It is easy to verify that the functions (6) and (7), as should be the case, satisfy the thermodynamic relation

$$P = \rho^2 \frac{dE}{d\rho},$$

because the specific entropy of matter with the equations of state (6) and (7) is zero. However, our subsequent calculations are seriously simplified if a simple representation of the derivatives of pressure in Eqs. (3) and (4) with (8) is used,

$$\frac{\partial P}{\partial \eta} = M\rho\frac{\partial \phi}{\partial \eta}, \quad \phi = \sqrt{1 + \xi_F^2} = \sqrt{1 + B^2\rho^{2/3}}, \tag{9}$$

and if the constant $M$ is expressed in terms of $B$ and $C$,

$$M = \frac{8}{3}CB^3 = 2.272 \times 10^{17} \text{ cm}^2 \text{ s}^{-2}.$$

We take the following simple analytic formula as the differential rotation law:

$$\Omega = \omega_0 \exp\left(-\frac{r^2 \sin^2\theta}{r_0^2}\right), \tag{10}$$

which is in satisfactory agreement with the collapse calculations by Imshennik and Nadyozhin (1992) if we substitute appropriate values for the arbitrary constant parameters $\omega_0$ and $r_0$ in it. Note that we may use any other plausible rotation law, for example, a law for which the rotation frequency decreases with increasing cylindrical radius as $1/\tilde{\omega}^k$. However, because of the stringent constraint on the mass, virtually the same toroidal solutions with slightly different profiles are obtained in all cases.

Using formula (9), let us write Eqs. (3) and (4) in final form, which we will solve to determine the structure of a hydrostatically equilibrium iron atmosphere in the form of $\rho = \rho(r, \theta)$ for the rotation law (10):

$$M\frac{\partial \phi}{\partial r} = r\omega_0^2 \exp\left(-\frac{2r^2\sin^2\theta}{r_0^2}\right)\sin^2\theta - \frac{GM_p}{r^2}, \tag{11}$$

$$M\frac{\partial \phi}{\partial \theta} = r^2\omega_0^2 \exp\left(-\frac{2r^2\sin^2\theta}{r_0^2}\right)\sin\theta\cos\theta, \tag{12}$$

where $\phi$ is defined in (9).

**3.** The system of equations (11) and (12) can be easily solved. Let us first integrate (11):

$$M\phi = -\frac{\omega_0^2 r_0^2}{4}\exp\left(-\frac{2r^2\sin^2\theta}{r_0^2}\right) + \frac{GM_p}{r} + F(\theta), \tag{13}$$

where $F(\theta)$ is an arbitrary function of $\theta$. In order to determine $F(\theta)$, we differentiate (13) with respect to $\theta$ and substitute the result in (12):

$$F'(\theta) = 0, \quad F(\theta) = F_0 = \mathrm{const}. \tag{14}$$

Thus, the sought-for solution is

$$M\phi = -\frac{\omega_0^2 r_0^2}{4}\exp\left(-\frac{2r^2\sin^2\theta}{r_0^2}\right) + \frac{GM_p}{r} + F_0; \tag{15}$$

hence, we can easily find $\rho = \rho(r, \theta)$ by using the definition of $\phi$ from (9):

$$\rho = \rho(r, \theta) = \frac{1}{B^3}\left\{\frac{1}{M^2}\left[\frac{GM_p}{r}\right.\right.$$

$$\left.\left. - \frac{\omega_0^2 r_0^2}{4}\exp\left(-\frac{2r^2\sin^2\theta}{r_0^2}\right) + F_0\right]^2 - 1\right\}^{3/2}. \tag{16}$$

Equation (16) contains an arbitrary integration constant of the system (11) and (12), which is yet to be determined from the boundary conditions. It is important to note that the system (11) and (12) has a solution only if we take $f(\theta) = \sin\theta$ in $\Omega = \Omega(rf(\theta))$; no solution exists for any other function $f(\theta)$, as would be expected (Tassoul 1978). In this case, the form of dependence on the argument $r\sin\theta$ is unimportant. We chose it in the form (10).

Next, we consider the boundary conditions to determine the constant $F_0$. Let the external radius of the atmosphere at which the density takes on a constant value $\rho_0$ be equal to $r_{ex}$. Note that it would be more natural to specify the boundary condition at the equator of the atmosphere, i.e., at $\theta_0 = \pi/2$, though relation (16) formally requires such a boundary condition only at some unique angle $\theta_0$. After simple transformations, we derive the final expression for the density distribution of an iron atmosphere by eliminating the integration constant $F_0$ from (16) using the above boundary condition:

$$\rho = \frac{1}{B^3}\left\{\left[\sqrt{1+B^2\rho_0^{2/3}} + \frac{GM_p}{M}\left(\frac{1}{r}-\frac{1}{r_{ex}}\right)\right.\right.$$
$$\left.\left.+ \frac{\omega_0^2 r_0^2}{4M}\left(\exp\left(-\frac{2r_{ex}^2}{r_0^2}\right) - \exp\left(-\frac{2r^2\sin^2\theta}{r_0^2}\right)\right)\right]^2 - 1\right\}^{3/2}. \quad (17)$$

As can be verified, only the factor $\sin^2\theta_0$ appears in the first exponential term in (17) at other angles $\theta_0 \neq \pi/2$. However, choosing the boundary condition not at the equator for constant $r_{ex}$ (particularly at small $\theta_0$) generally causes the total atmospheric mass to increase considerably, which seems unnatural.

**4.** Let us write some integrated quantities that are of interest in physically interpreting the solution (17) with allowance for a mirror symmetry. The total atmospheric mass $M_0$ for the given density distribution $\rho(r, \theta)$ (17) is

$$M_0 = 2 \times 2\pi \int_{r_{in}}^{r_{ex}} r^2 dr \int_0^{\pi/2} \rho(r, \theta)\sin\theta d\theta. \quad (18)$$

The total angular momentum $J_0$ of the atmosphere is

$$J_0 = 2 \times 2\pi \int_{r_{in}}^{r_{ex}} r^4 dr \int_0^{\pi/2} \rho(r,\theta)\Omega(r\sin\theta)\sin^3\theta d\theta. \quad (19)$$

The gravitational ($E_{gr}$), internal ($E_{in}$) [using (7) for $E(\rho)$], and rotational ($E_{rot}$) energies are, respectively,

$$E_{gr} = -2 \times 2\pi \int_{r_{in}}^{r_{ex}}\left(\frac{GM_p}{r}\right)r^2 dr \int_0^{\pi/2}\rho(r,\theta)\sin\theta d\theta, \quad (20)$$

$$E_{in} = 2 \times 2\pi \int_{r_{in}}^{r_{ex}} r^2 dr \int_0^{\pi/2}\rho(r,\theta)E(\rho)\sin\theta d\theta, \quad (21)$$

$$E_{rot} = 2 \times 2\pi \int_{r_{in}}^{r_{ex}} r^4 dr$$
$$\times \int_0^{\pi/2}\rho(r,\theta)\left(\frac{1}{2}\Omega^2(r\sin\theta)\right)\sin^3\theta d\theta. \quad (22)$$

All integrals (18)–(22) are written in this case of symmetry about the $z = 0$ equatorial plane; they imply the presence of a matter-free spatial region for which the expression in curly braces in (17) becomes a negative quantity without any physical meaning. Note that the equilibrium equations (11) and (12) formally admit of an imaginary solution $\rho$, though the function $\phi$ [see relation (9)] in Eqs. (11) and (12) is, of course, always positive. Therefore, the domain of the solution of interest to us is bounded by the line of zero density $\rho = 0$ (17):

$$f(r_{in}^*, \theta) = \sqrt{1+B^2\rho_0^{2/3}} + \frac{GM_p}{M}\left(\frac{1}{r_{in}^*}-\frac{1}{r_{ex}}\right)$$
$$+ \frac{\omega_0^2 r_0^2}{4M}\left[\exp\left(-\frac{2r_{ex}^2}{r_0^2}\right) - \exp\left(-\frac{2r_{in}^{*2}\sin^2\theta}{r_0^2}\right)\right] - 1 = 0. \quad (23)$$

As our numerical solution of the transcendental equation (23) shows, the angle $\theta$ at small values of the product $r_0\omega_0 (\leq 1.08 \times 10^{11}$ cm s$^{-1}$) varies in the range from $\theta = \theta_{max} = \pi/2$ to $\theta = \theta_{ex} < \pi/2$ while passing through a minimum $\theta = \theta_{min} < \theta_{ex}$. At large values of $r_0\omega_0$, the angle $\theta$ monotonically decreases from $\pi/2$ to $\theta_{ex}$. In all cases, the radius $r_{in}^*$ monotonically increases from its minimum $r_{in\,min}^* \equiv r_{min}^*$ to maximum $r_{in\,max}^* \equiv r_{ex}$. This is how the inner surface of the toroidal volume of an iron atmosphere is obtained, whereas its outer surface coincides with the external sphere of radius $r = r_{ex}$, and the angle clearly increases from $\theta = \theta_{ex}$ to $\theta = \pi/2$. In this case, $\rho = \rho_0$ at the equator by our definition of the constant $F_0$ described above when passing from (16) to (17); $\rho = 0$ at $\theta = \theta_{ex}$, and, using (23), we explicitly derive at $r_{in}^* = r_{ex}$

$$\exp\left(-\frac{2r_{ex}^2\sin^2\theta_{ex}}{r_0^2}\right) = \frac{4M}{\omega_0^2 r_0^2}\sqrt{1+B^2\rho^{2/3}}$$
$$+ \exp\left(-\frac{2r_{ex}^2}{r_0^2}\right) - \frac{4M}{\omega_0^2 r_0^2}. \quad (24)$$

Here, we are primarily interested in low-mass atmospheres with $M_0 \simeq 0.1 M_\odot$. As our calculations show, they are toroidal structures with $r_{ex} > r_{min}^* > r_0 \gg r_{in}$ and $\theta_{min} \approx \theta_{ex} \sim \pi/4$. Therefore, $r_{in}$, which is such an uncertain quantity for a NS binary with a total mass $M_p + M_{ns} = M_{Fe} - M_e$, is of no practical importance.

**5.** Next, it is instructive to show how the solution changes when we pass from the general equation of state (6) to a polytropic one; i.e., when the dependence $P = P(\rho)$ takes the form

$$P = K\rho^\gamma, \quad (25)$$

where the coefficient $K$ can be formally represented as $K = P_0/\rho_0^\gamma$. This passage affects only the form of the dependence $\phi = \phi(\rho)$. When the equation of state (6) is

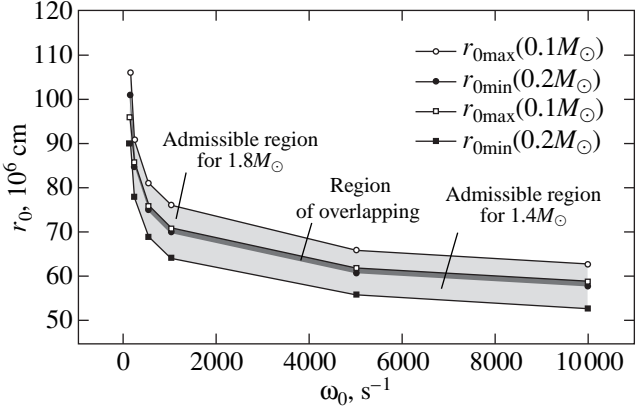

**Fig. 1.** Admissible (by the atmospheric mass) regions in the $r_0$–$\omega_0$ plane for central masses $M_p = 1.8$ and $1.4 M_\odot$.

replaced by expression (25), the function $\phi$, according to (9), takes the form

$$M\phi = \frac{K\gamma}{\gamma-1}\rho^{\gamma-1}. \tag{26}$$

Thus, since the entire line of reasoning when deriving $\phi = \phi(r, \theta)$ from the solution of Eqs. (11) and (12) does not depend on the form of the equation of state and is valid as before, we can immediately write the expression for the density distribution of an iron atmosphere in the polytropic approximation:

$$\rho = \rho(r, \theta) = \left[\rho_0^{\gamma-1} + \frac{GM_p}{A(\gamma)}\left(\frac{1}{r} - \frac{1}{r_{ex}}\right)\right.$$
$$\left. + \frac{\omega_0^2 r_0^2}{4A(\gamma)}\left(\exp\left(-\frac{2r_{ex}^2}{r_0^2}\right) - \exp\left(-\frac{2r^2 \sin^2\theta}{r_0^2}\right)\right)\right]^{\frac{1}{\gamma-1}}, \tag{27}$$

where $A(\gamma) = K\gamma/\gamma - 1$. Below, we also give explicit expressions for the density functions $\rho(r, \theta)$ from (27) in two limiting cases for the equation of state (6):

$$\xi_F \ll 1: \gamma = \frac{5}{3},$$

$$K_{5/3} = \frac{8}{15}CB^5 = 2.758 \times 10^{12} \text{ cm}^4 \text{ s}^{-2} \text{ g}^{-2/3},$$

$$\rho = \left[\rho_0^{2/3} + \frac{2GM_p}{5K_{5/3}}\left(\frac{1}{r} - \frac{1}{r_{ex}}\right)\right. \tag{28}$$
$$\left. + \frac{\omega_0^2 r_0^2}{10K_{5/3}}\left(\exp\left(-\frac{2r_{ex}^2}{r_0^2}\right) - \exp\left(-\frac{2r^2 \sin^2\theta}{r_0^2}\right)\right)\right]^{\frac{3}{2}},$$

$$\xi_F \gg 1: \gamma = \frac{4}{3},$$

$$K_{4/3} = \frac{2}{3}CB^4 = 4.424 \times 10^{14} \text{ cm}^3 \text{ s}^{-2} \text{ g}^{-1/3},$$

$$\rho = \left[\rho_0^{1/3} + \frac{GM_p}{4K_{4/3}}\left(\frac{1}{r} - \frac{1}{r_{ex}}\right)\right. \tag{29}$$
$$\left. + \frac{\omega_0^2 r_0^2}{16K_{4/3}}\left(\exp\left(-\frac{2r_{ex}^2}{r_0^2}\right) - \exp\left(-\frac{2r^2 \sin^2\theta}{r_0^2}\right)\right)\right]^3,$$

where $B$ and $C$ are the known constants. Note that, for a given rotation law, replacing the general equation of state by a polytropic one, as our calculations show, causes no appreciable change in the shape of the atmospheric surface, but causes a substantial density redistribution inside the atmosphere and a considerable reduction in mass (particularly for $\gamma = 5/3$).

**6.** We now turn to the results of our numerical calculations of density distributions $\rho(r, \theta)$ [relation (17)] and of the corresponding integrated quantities (18)–(22). In order for the validity conditions of the Roche approximation and the conditions of gravitational collapse with a residual iron gas mass $M_e$ (see above) of interest to us to be satisfied, we restrict the total mass of an equilibrium atmosphere $M_0$ (18) to the range (0.1–0.2)$M_\odot$. We take the external radius to be $r_{ex} = 2.38 \times 10^8$ cm and the limiting equatorial density to be $\rho_0 = 5.66 \times 10^5$ g cm$^{-3}$ (Imshennik and Zabrodina 1999). Consider two masses of the central NS: $M_p = 1.8 M_\odot$ and $1.4 M_\odot$ (see above). Our numerical calculations indicate that a fairly narrow admissible (by the total mass $M_0$ of an iron atmosphere) region in the $r_0 - \omega_0$ plane is obtained for a fixed mass of the central NS. Figure 1 shows two such regions for the two $M_p$, which slightly overlap with one another. We see from Fig. 1 that, as the central mass decreases, the admissible region displaces to the coordinate origin in the $r_0 - \omega_0$ plane, i.e., toward atmospheres with slower rotation (with a smaller product $r_0\omega_0$), as should be the case. The table gives the rotation parameters of these regions; quite naturally, $r_{0\max}$ and $r_{0\min}$ for a given $\omega_0$ correspond to the minimum and maximum admissible atmospheric masses $M_0 = 0.1$ and $0.2 M_\odot$, respectively (see also Fig. 1). Clearly, for any given $M_p$, more massive atmospheres with $M_0 > 0.2 M_\odot$ and less massive atmospheres with $M_0 < 0.1 M_\odot$ lie, respectively, below and above the highlighted regions in Fig. 1. The table also lists values for all forms of energy (20)–(22), as well as the total energy $E_{tot} = E_{gr} + E_{in} + E_{rot}$ and total angular momentum $J_0$ (19). All these quantities refer to both boundaries of the regions inside which they continuously change from one limiting value to the other. We see from the table that all $E_{tot} < 0$, with $|E_{tot}| \sim 10^{50}$ erg, for all $\omega_0$ except the slowest rotation $\omega_0 = 100$ s$^{-1}$. In this case, $|E_{tot}|$ sharply

**Table**

| | $M = 1.8M_0$, $r_{ex} = 2.38 \times 10^8$ cm | | | | | | |
|---|---|---|---|---|---|---|---|
| $\omega_0$, s$^{-1}$ | | 100 | 200 | 500 | 1000 | 5000 | 10000 |
| $r_{0\,max}$, $10^6$ cm | | 106 | 91 | 81 | 76 | 66 | 63 |
| $r_{0\,min}$, $10^6$ cm | | 101 | 85 | 75 | 70 | 61 | 58 |
| $J_{tot\,max\,r_0}$ | , $10^{49}$ cm$^2$ s$^{-1}$ g | 2.07 | 3.18 | 3.09 | 2.8 | 2.67 | 2.54 |
| $J_{tot\,min\,r_0}$ | | 3.41 | 5.27 | 5.17 | 4.79 | 4.3 | 4.15 |
| $E_{gr\,max\,r_0}$ | , $10^{50}$ erg | −28.9 | −2.73 | −2.79 | −2.54 | −2.62 | −2.55 |
| $E_{gr\,min\,r_0}$ | | −69.7 | −5.38 | −5.64 | −5.35 | −5.17 | −5.14 |
| $E_{rot\,max\,r_0}$ | , $10^{50}$ erg | 0.78 | 1.11 | 1.12 | 1.02 | 1.05 | 1.02 |
| $E_{rot\,min\,r_0}$ | | 1.49 | 2.08 | 2.16 | 2.04 | 1.96 | 1.95 |
| $E_{in\,max\,r_0}$ | , $10^{50}$ erg | 5.10 | 0.84 | 0.87 | 0.80 | 0.84 | 0.82 |
| $E_{in\,min\,r_0}$ | | 15.0 | 1.79 | 1.89 | 1.80 | 1.77 | 1.77 |
| $E_{tot\,max\,r_0}$ | , $10^{50}$ erg | −23.1 | −0.79 | −0.80 | −0.72 | −0.74 | −0.72 |
| $E_{tot\,min\,r_0}$ | | −53.2 | −1.54 | −1.59 | −1.50 | −1.44 | −1.43 |
| | $M = 1.4M_0$, $r_{ex} = 2.38 \times 10^8$ cm | | | | | | |
| $\omega_0$, s$^{-1}$ | | 100 | 200 | 500 | 100 | 5000 | 10000 |
| $r_{0\,max}$, $10^6$ cm | | 96 | 86 | 76 | 71 | 62 | 59 |
| $r_{0\,min}$, $10^6$ cm | | 90 | 78 | 69 | 64 | 56 | 53 |
| $J_{tot\,max\,r_0}$ | , $10^{49}$ cm$^2$ s$^{-1}$ g | 2.47 | 2.44 | 2.45 | 2.29 | 2.08 | 2.01 |
| $J_{tot\,min\,r_0}$ | | 3.76 | 4.18 | 3.94 | 3.75 | 3.27 | 3.19 |
| $E_{gr\,max\,r_0}$ | , $10^{50}$ erg | −7.65 | −2.08 | −2.25 | −2.16 | −2.12 | −2.11 |
| $E_{gr\,min\,r_0}$ | | −25.0 | −4.55 | −4.60 | −4.57 | −4.29 | −4.39 |
| $E_{rot\,max\,r_0}$ | , $10^{50}$ erg | 0.68 | 0.81 | 0.86 | 0.83 | 0.81 | 0.81 |
| $E_{rot\,min\,r_0}$ | | 1.49 | 1.66 | 1.66 | 1.64 | 1.54 | 1.56 |
| $E_{in\,max\,r_0}$ | , $10^{50}$ erg | 1.56 | 0.77 | 0.85 | 0.83 | 0.82 | 0.82 |
| $E_{in\,min\,r_0}$ | | 5.76 | 1.79 | 1.85 | 1.85 | 1.76 | 1.81 |
| $E_{tot\,max\,r_0}$ | , $10^{50}$ erg | −5.30 | −0.50 | −0.53 | −0.51 | −0.49 | −0.49 |
| $E_{tot\,min\,r_0}$ | | −17.8 | −1.09 | −1.09 | −1.08 | −1.0 | −1.02 |

increases (by several tens of times). Clearly, this is of no interest from the viewpoint of the model for asymmetric explosions of collapsing supernovae (Aksenov et al. 1997), because the atmospheres closely approach the central star, up to the contact ($r_{min}^* = r_{in}$), and become gravitationally bound in explosion ($|E_{gr}| \sim 10^{52}$ erg). At the same time, for all other values of $\omega_0 \geq 200$ s$^{-1}$, the energy parameters of the atmospheres are acceptable in the above ratio: the contributions of $E_{in}$ and $E_{rot}$ turn out to be similar, $\sim 10^{50}$ erg, while $|E_{gr}| \sim 3 \times 10^{50}$ erg. Note also that the table gives the corresponding values of total angular momentum $J_0$, $(2-4) \times 10^{49}$ erg s; they vary in a relatively narrow range consistent with the calculations of Imshennik and Nadyozhin (1992). Figures 2–4 show isochores in the $\theta$–$r$ plane of spatial variables, which demonstrate the configuration of a toroidal atmosphere for its density distribution $\rho(r, \theta)$ for the general equation of state at three different points of the admissible region for the central mass $M_p = 1.8M_\odot$. The

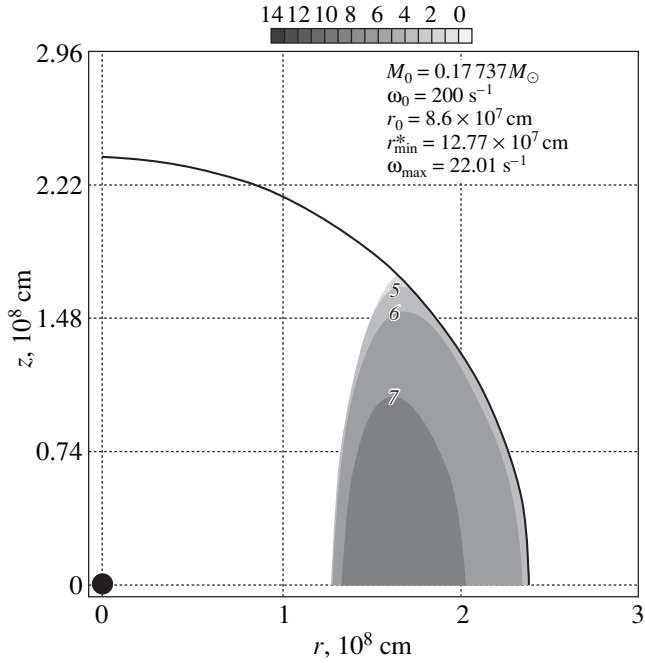

**Fig. 2.** Isochores in the θ–r plane of spatial variables showing the atmospheric density distribution ρ(r, θ) for the general equation of state with a central mass $M_p = 1.8 M_\odot$ and for rotation parameters $r_0 = 86 \times 10^6$ cm and $\omega_0 = 200$ s$^{-1}$.

figures also show minimum radii $r^*_{min}$ for the inner boundary of the toroidal volume of an iron atmosphere and maximum rotation frequencies $\omega_{max}$ of the atmospheric matter. This is clearly reached just near the inner boundary with radius $r^*_{min}$ [see (10)]. These data lead us to conclude that, at all points within the admissible (by the atmospheric mass) region shown in Fig. 1, we obtain toroidal solutions with a characteristic highly elongated (along the z axis) profile differing only slightly from each other. The characteristic features shared by all solutions are a maximum matter density in the atmosphere ($10^7$–$10^8$ g cm$^{-3}$) and an abrupt drop in density at the inner boundary. When the mass of the central NS is $M_p = 1.4 M_\odot$, we obtain almost the same results as would be expected. For comparison, Figs. 5 and 6 show the profiles of atmospheres with the same rotation law as in Fig. 3, but for a polytropic equation of state. Figures 5 and 6 correspond to atmospheres with nonrelativistic (polytropic index γ = 5/3) and relativistic (γ = 4/3) equations of state, respectively. Both cases are characterized by an appreciable reduction in total atmospheric mass compared to the general equation of state through a slight change in the internal structure of the atmosphere, with the shape of the outer boundary being virtually unchanged. This effect turns out to be particularly strong for atmospheres with the polytropic index γ = 5/3.

7. Let us discuss our results for toroidal iron atmospheres with rotation in the vicinity of a central star located near the coordinate origin with the mass of the more massive NS from the putative NS binary. Of fundamental importance is the very existence of steady-state solutions for such atmospheres with acceptable sets of parameters for an asymmetric explosion with a typical supernova energy to take place. The above statement of the problem, which assumes a zero matter tem-perature, lifts any upper bounds on the lifetimes of these atmospheres. At the same time, it is clear that the assumption of a zero temperature requires higher rotation parameters $\omega_0$ and $r_0$ than those in the general case of nonzero temperatures. It is instructive to roughly estimate the time scale of cooling through neutrino energy losses for the matter of an iron atmosphere with density $\rho \simeq 10^6$ g cm$^{-3}$. Assuming a kinetic equilibrium for beta processes with volume emission of pairs of electron neutrinos ($\nu_e$ and $\tilde{\nu}_e$), we may take (Imshennik et al. 1967) $q_{\nu,\tilde{\nu}}^{(n,p)} \simeq 10^{18}$ erg cm$^{-3}$ s$^{-1}$ for the power of energy losses in matter composed of free nucleons at temperature $T = 2 \times 10^9$ K (see Fig. 1 in the above paper). In our case of matter composed of completely ionized iron, this estimate, roughly speaking, must be reduced by a factor of $A_{Fe}(ft)_{Fe}/(ft)_{n,p}$, where $A_{Fe} = 56$, $(ft)_{Fe} \simeq 10^5$ and $(ft)_{n,p} \simeq 10^3$; i.e., $q_{\nu,\tilde{\nu}}^{(Fe)} \simeq 2 \times 10^{15}$ erg cm$^{-3}$ s$^{-1}$. If we write the cooling time scale as $\tau_q \simeq e_{in}/q_{\nu,\tilde{\nu}}^{(Fe)}$, where the internal energy density is

$$e_{in} = \frac{3}{2}k_B T \frac{(Z_{Fe}+1)\rho}{A_{Fe}m_0} \simeq 1.2 \times 10^{23} \frac{\text{erg}}{\text{cm}^3},$$

we then obtain the final estimate $\tau_q \simeq 6 \times 10^7$ s $\simeq 2$ years. This estimate essentially corresponds to the conditions of a nondegenerate iron gas and neglect of all energy thresholds for beta processes; i.e., at least $k_B T \sim m_e c^2 \sim E_0 \sim 1$ MeV and $T \geq T_d \simeq 2 \times 10^9$ K ($E_0$ is the characteristic threshold energy of electron beta capture in iron, and $T_d$ is the degeneracy temperature). Our rough estimate of $\tau_q$ implies that, over the lifetimes of the NS binaries (~1 h) of interest to us, no significant matter cooling in their atmospheres takes place. Therefore, the atmospheric structures found in our analytic solution slightly overestimate the rotation parameters $\omega_0$ and $r_0$ required for hydrostatic equilibrium. Finally, note that the above temperature $T \simeq 2 \times 10^9$ K just corresponds to the results of our collapse calculations (Imshennik and Nadyozhin 1992; see also Imshennik and Zabrodina 1999). The complex problem of the behavior of the outer layers in SN progenitors, in particular, in circumpolar vacuum holes, needs further study.

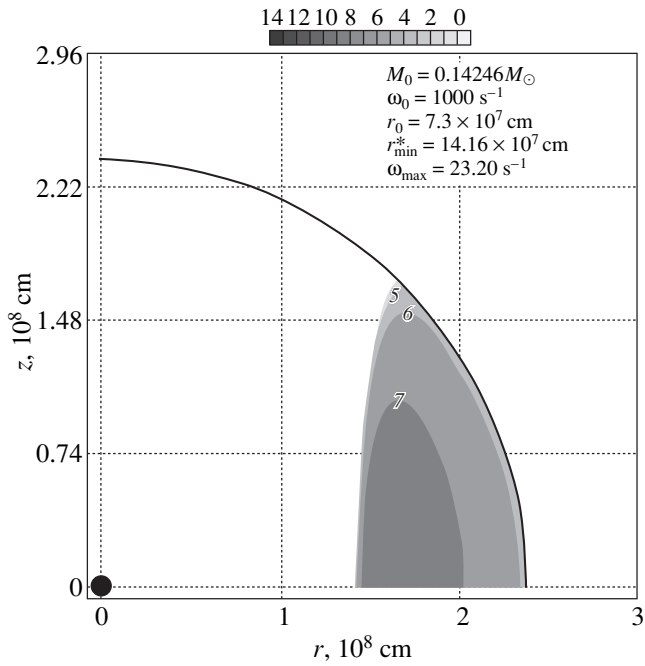

**Fig. 3.** Same as Fig. 2 for $r_0 = 73 \times 10^6$ cm and $\omega_0 = 1000$ s$^{-1}$.

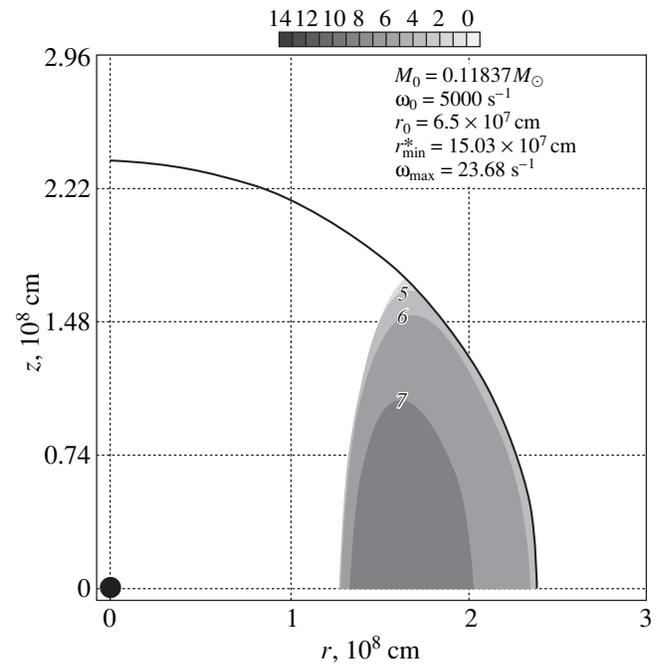

**Fig. 4.** Same as Fig. 2 for $r_0 = 65 \times 10^6$ cm and $\omega_0 = 5000$ s$^{-1}$.

Having expanded in the transverse direction, the explosion remnants of a low-mass NS moving at a high velocity in an equatorial circular orbit eventually reach the inner surface of such toroidal atmospheres and produce shock waves inside them. Note, for completeness, that the rotation axes of the binaries and the atmospheres coincide, being the rotation axes of the original stellar iron cores. Thus, shock fronts propagate from the centers of the atmospheres to their periphery, which seem to gradually acquire the characteristic features established in our previous two-dimensional hydrodynamic calculations of asymmetric explosions (Aksenov et al. 1997; Imshennik and Zabrodina 1999; Zabrodina and Imshennik 2000). Nevertheless, the new problems of shock generation and propagation in toroidal atmospheres differ significantly because of the nonuniform density distributions inside the atmospheres. However, since the mean densities of these atmospheres are approximately equal to the constant density $\rho_0 = 5.66 \times 10^5$ g cm$^{-3}$ assumed in our previous calculations, such differences will vanish in the course of time. In addition, the dependence of the hydrodynamic pattern on the preshock matter density is generally weak. In particular, the previously found high degree of asymmetry for late times must be the same in the new statement of the problem. Next, in all probability, the iron dissociation in toroidal atmospheres is considerably reduced because of the relatively high matter density near the inner surface of these atmospheres. After shock generation, the circumpolar vacuum cavities must be partly filled with the matter that passed through the shock front. We conclude that new calculations of asymmetric explosions with allowance for the steady-state toroidal atmospheres obtained here are undoubtedly of considerable interest. However, it would apparently be unreasonable to expect radical changes in the hydrodynamic pattern of explosion compared to our previous calculations in the approximation of homogeneous atmospheres.

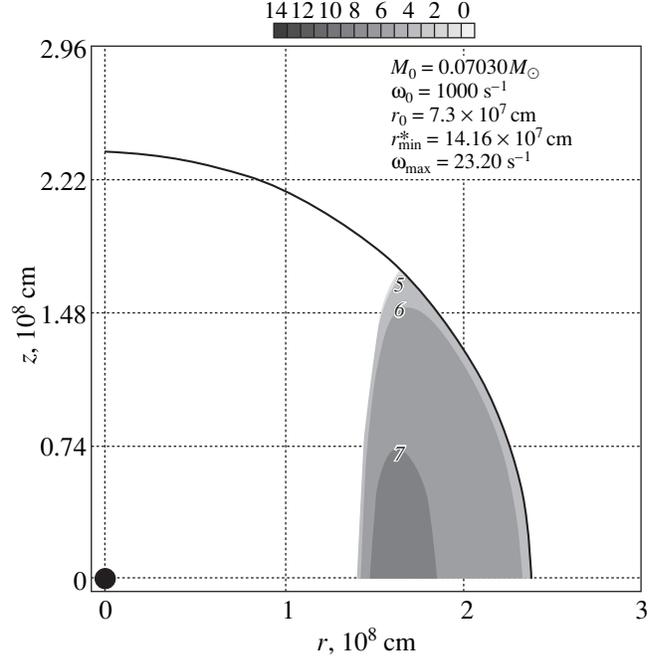

**Fig. 5.** Isochores in the $\theta$–$r$ plane of spatial variables showing the atmospheric density distribution $\rho(r, \theta)$ for a polytropic equation of state with $\gamma = 5/3$, a central mass $M_p = 1.8 M_\odot$, and rotation parameters $r_0 = 73 \times 10^6$ cm and $\omega_0 = 1000$ s$^{-1}$.

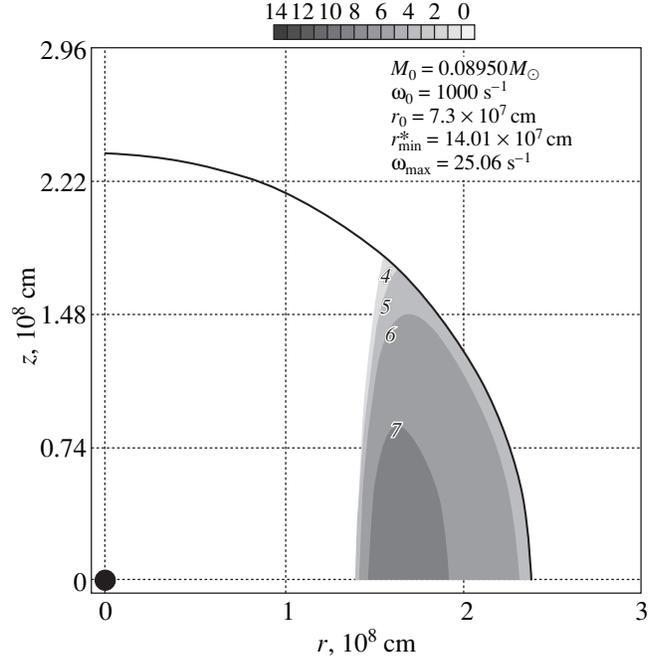

**Fig. 6.** Same as Fig. 5 for $\gamma = 4/3$.

## ACKNOWLEDGMENTS

We wish to thank M.S. Popov for taking an active part in the discussion of this study. One of us (V.S. Imshennik) is also grateful to Prof. Hillebrandt, who emphasized the importance of the existence of steady-state atmospheres in binaries of neutron stars for the explosion scenario of collapsing supernovae under consideration.

*Translated by V. Astakhov*